\documentclass[aps, pra, twocolumn,superscriptaddress,footinbib]{revtex4}

\usepackage{amsmath,amssymb}
\usepackage{ctable}
\input{epsf}
\input{epsfx}

\def\ket#1{\mathinner{|{#1}\rangle}}
\def\braket#1{\mathinner{\langle{#1}\rangle}}

\usepackage{float}
\usepackage{graphicx}
\usepackage{amssymb}
\usepackage{ae} 
\usepackage{subfigure}
\usepackage{fmtcount}
\usepackage{upgreek}

\begin{document}

\title{Efficient generation of single and entangled photons on a silicon \\ photonic integrated chip}

\author{Jacob Mower}
\affiliation{Department of Electrical Engineering, Columbia University, New York, NY 10027}
\author{Dirk Englund}
\affiliation{Department of Electrical Engineering, Columbia University, New York, NY 10027}
\affiliation{Department of Applied Physics and Applied Mathematics, Columbia University, New York, NY 10027}

\begin{abstract}
We present a protocol for generating on-demand, indistinguishable single photons on a silicon photonic integrated chip. The source is a time-multiplexed spontaneous parametric down conversion element which allows optimization of single photon versus multiphoton emission while realizing high output rate and indistinguishability. We minimize both the scaling of active elements and the scaling of active element loss with multiplexing. We then discuss detection strategies and data processing to further optimize the procedure. We simulate an improvement in single photon generation efficiency over previous time-multiplexing protocols, assuming existing fabrication capabilities.  We then apply this system to generate heralded Bell states. The generation efficiency of both nonclassical states could be increased substantially with improved fabrication procedures.

~
\end{abstract}

\maketitle

\section{Introduction}

Single photon sources are central to a number of key experiments in quantum information science, including tests of quantum nonlocality \cite{PhysRevLett.47.460, Merali18032011}, quantum key distribution \cite{springerlink:10.1007/BF00191318,PhysRevLett.67.661}, and quantum teleportation \cite{PhysRevLett.70.1895, Bouwmeester97nature, pan_memory_2008}. Complex non-classical optical states required for quantum metrology can be constructed from suitable single photon sources \cite{PhysRevA.65.052104}, and recent theoretical proposals have shown that nondeterministic logic operations between multiple photonic qubits, combined with the ability to detect when gates have succeeded (feed-forwardability), allows for efficient quantum computation \cite{klm_2001, PhysRevLett.93.040503}. To realize these and other quantum information technologies, a number of experimental efforts are underway to produce efficient sources of indistinguishable single photons, including the application of quantum dots in micro- and nanocavities \cite{Abram2005APL, strauf_high_freq_2007, 2007.OpEx.Englund, PhysRevLett.104.137401, gerard_qd_nanowire_2010}, isolated cold atoms \cite{2007.Nature.Rempe.single_photon}, and isolated single molecules in solid state systems \cite{PhysRevLett.104.123605}. However these experiments have not realized photon indistinguishability as high as that from spontaneous parametric down conversion \cite{PhysRevA.60.R773}, and require complex setups including high vacuum and cryogenic temperatures not immediately suitable for scalability.

In this work, we present a scheme to integrate a bright, efficient source of highly indistinguishable photons on a silicon-on-insulator (SOI) photonic integrated circuit (PIC). We call this scheme actively multiplexed parametric photon (AMPP) generation. Compared to bulk optics, PICs offer the advantage of miniaturization, high field intensity, phase stability, high mode overlap in coupling regions \cite{Politi02052008} and the capacity to cheaply increase component number. It has recently been shown that PIC platforms are also capable of implementing high fidelity multiqubit operations \cite{laing:211109, 5340091}. 

While quantum optics experiments on PICs have been performed using low-index contrast oxide systems, entanglement experiments on the SOI platform have not to our knowledge been reported. SOI chips have a number of advantages, leveraging advanced fabrication for scalability, electro-optic switching and on-chip integration of state-of-the-art electronics. SOI chips for classical signal processing are also being advanced for applications such as optical interconnects for high-performance computing systems (cf. Ref. \cite{vlasov}).

In Section II, we present our protocol for the efficient generation of single photons at $1560$ nm using an optical circuit integrated on the SOI platform. In Section III we discuss the optimization of system parameters to maximize efficiency given realistic operating conditions. We apply this optimized system in Section IV to generate heralded Bell states. We frame this paper in the context of SOI networks. The same design principles can be applied to other integrated systems, as well as bulk optical systems.

\section{AMPP generation protocol}

The AMPP source uses photon pairs generated by spontaneous parametric down conversion (SPDC) \cite{PhysRevLett.61.2921}. SPDC is a second-order nonlinear optical process characterized by the interaction Hamiltonian, $\hat{H}_{int}=i\chi\hbar(a_{s}^{\dagger}a_{i}^{\dagger}-a_{s}a_{i})$, where $a_{s}$ ($a_{i}$) is the annihilation operator corresponding to the signal (idler) photon and $\chi=E\chi^{(2)}$, where $\chi^{(2)}$ is the second order nonlinear susceptibility tensor and $E$ is a classical pump field \cite{walls_milburn}. The time-evolved state of the signal and idler photons is given by

\begin{equation}
	|\psi(t)\rangle=e^{-i\hat{H}t/\hbar}|0,0\rangle=e^{-\chi t(a_{s}^{\dagger}a_{i}^{\dagger}-a_{s}a_{i})}|0,0\rangle
\end{equation}

This can be expanded in the Fock basis to calculate the probability of generating $n$ pairs in some time $t$, $P_{n}=|\langle n,n|\psi(t)\rangle|^{2}\approx (n+1)(\lambda/2)^n e^{-\lambda}$, where $|n,m\rangle$ represents $n$ ($m$) photons in the signal (idler) rail \cite{PhysRevA.61.042304} and $\lambda=2\tanh^2 \chi t$. Therefore as $\lambda\rightarrow0$, $|\psi(t)\rangle\rightarrow|0,0\rangle+\lambda|1,1\rangle$, a single pair state. Detecting the idler photon of each pair indicates the existence of the signal photon \cite{PhysRevLett.56.58}, which yields highly indistinguishable \cite{PhysRevA.60.R773} heralded single photons. In this paper, we will assume the use of periodically poled potassium tytanil phosphate (PPKTP) waveguide sources, which have been shown to produce highly indistinguishable photon pairs \cite{Zhong:09}, and to enable efficient heralded single photon emission \cite{Levine:10}. Unfortunately, due to the approximately Poissonian number state distribution, the source must be driven weakly so that $P_1 \ll 1$. Such a source is not suitable for many scalable quantum information technologies including linear optics quantum computation \cite{klm_2001}. 

It is possible to improve this efficiency by actively switching multiple sources into a single output, contingent upon heralding \cite{PhysRevA.66.053805,Shapiro:07,McCusker:08,PhysRevA.83.043814,PhysRevA.66.042303,1367-2630-6-1-100}. Migdall and collaborators first considered this approach using a set of $N$ distinct weakly-pumped SPDC crystals, switched into a single output by an $N$-by-$1$ switch \cite{PhysRevA.66.053805}. Later proposals using strictly $n\times m$ switches for $n,m \le 2$ require order of $N$ switch scaling \cite{Shapiro:07} to accomplish this spatial multiplexing. 

\begin{figure}[!t]
\centering\includegraphics[scale=0.41]{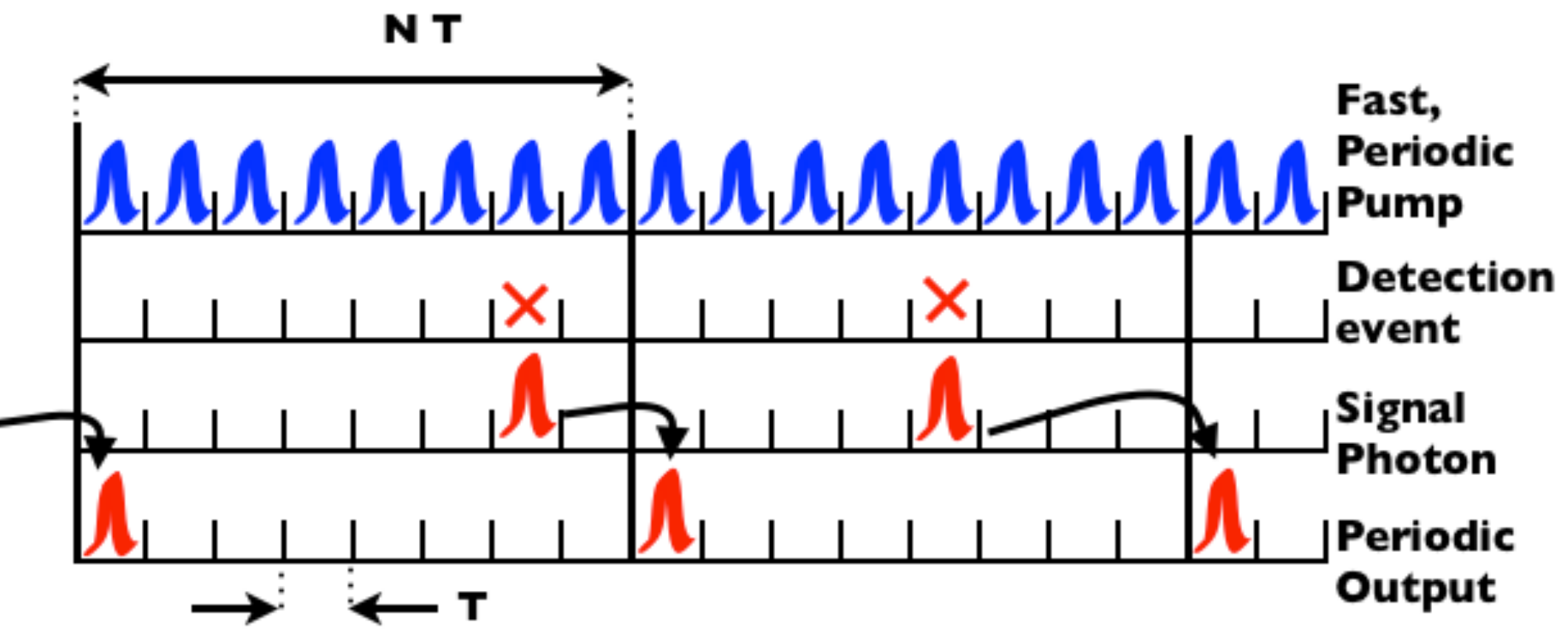}
\centering\caption{A nonlinear crystal is pumped at period $T$, resulting in the generation of photon pairs at random intervals. The idler photon is split off and detected, heralding the existence of the signal photon. The appropriate signal photon is then sent into a delay circuit, where it is delayed to the next rail at period $NT$.}
\label{fig:time}
\end{figure}

We are able to reduce the switch number to order of $\log_2 N$ by using the time-multiplexing scheme shown in Fig. \ref{fig:time}.  In this scheme a single SPDC element is pumped at some period $T$ so that photon pairs are generated according to the Poisson statistics in each time bin. The idler photons are sent to a detector and the corresponding signal photons routed to a variable delay circuit. Based on the detection of the idler photons, one signal photon is routed to a single time bin, while any others are rejected. Thus, the scheme targets pulsed single photon emission with a period of $NT$ \cite{PhysRevA.66.042303, McCusker:08}.

We alter the variable delay circuit of previous schemes from a single delay line to one composed of separate delays $j$ with time delay $2^j T$; any delay from $0$ to $T(N-1)$ can be constructed as $T\sum_{j=0}^{\lfloor \log_2 N\rfloor} c_j2^j$, where $c_j\in \{0,1\}$. In this representation, if $c_j=1$ ($=0$), then the photon is (not) routed into delay $j$. All delays are therefore achieved with order of $\log_2 N$ switches. Switching loss scales exponentially with the number of switches a photon passes through, so our switching loss scaling is of order $N$. Previous spatial multiplexing protocols had a switching loss that scaled as $N$, but required order of $N$ switching elements \cite{Shapiro:07,PhysRevA.83.043814}. The $\log_{2}N$ scaling of our protocol reduces energy consumption, increases resilience to fabrication imperfections, and eases scalability but it requires optical delay lines. Previous time-multiplexing protocols, optimized for implementation in bulk optics, require only one switching element. However for this design, we will show the signal photons pass this switch a greater number of times and therefore experience greater attenuation. This is especially important because switches represent the main source of loss in the on-chip implementation. 

\begin{figure*}
	\centering\includegraphics[scale=0.42]{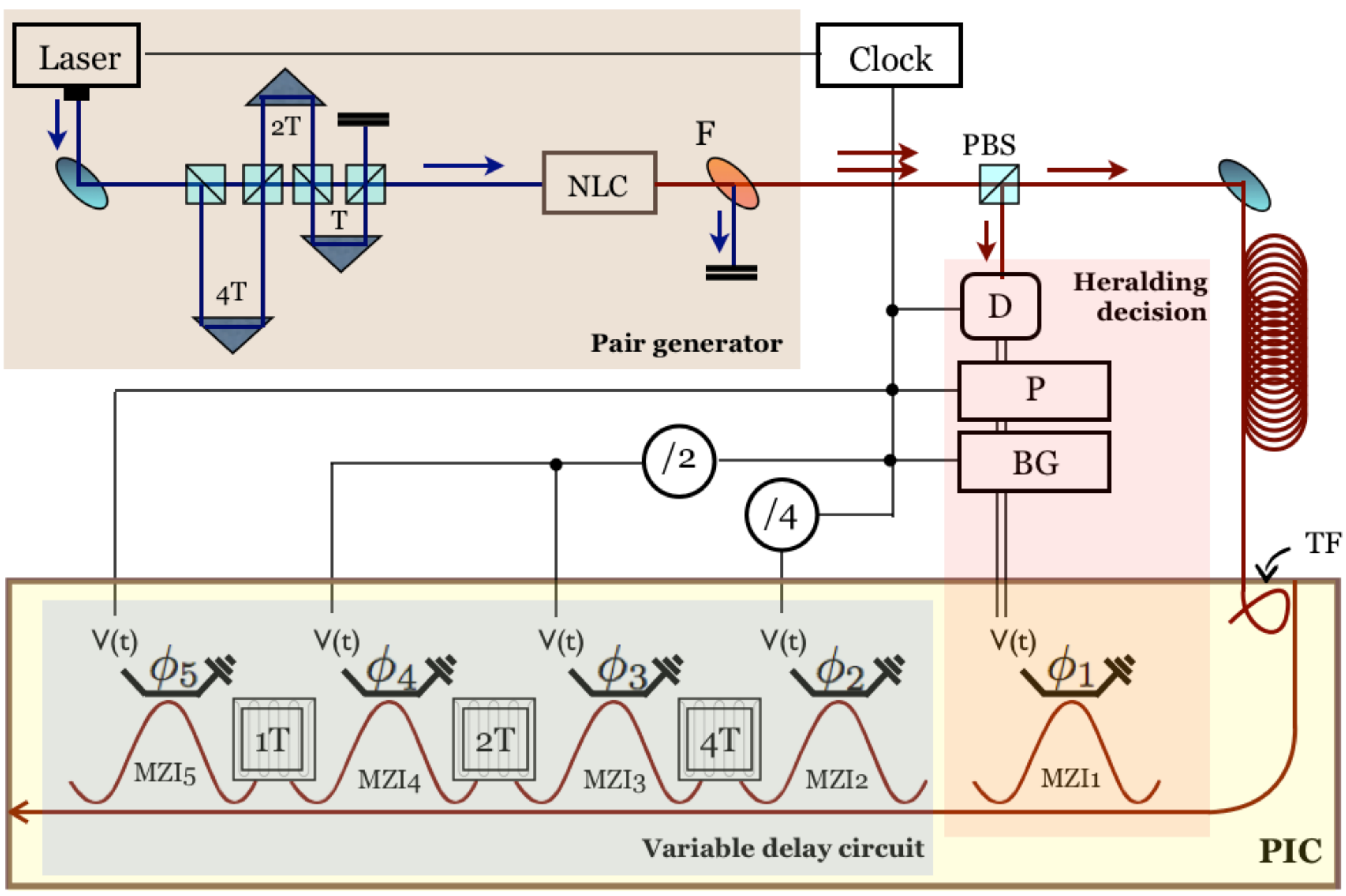}
	\caption{AMPP generation for $N=8$. A laser enters a series of beam splitters and delay arms designed to generate eight pulses at period $T$. The laser then pumps a nonlinear crystal (NLC) and emits an electronic triggering signal, which gates the heralding detector (D) and synchronizes the decision electronics. The photon pairs generated are split at a polarizing beam splitter (PBS), with the idler photon sent to the detector. The detector outputs a bit stream which is read by the data processor (P). The processor outputs a bit stream to an ultra-fast bit generator, which controls the switch, MZI1. The signal photon is sent to a delay line, and is either rejected by MZI1 or transmitted to the active delay circuit. MZI2 - 5 are controlled by periodic clock cycles. $/n$ denotes the clock rate division by a factor of $n$ required to drive the modulators according to Table \ref{tab:logic}. Delays are denoted by $nT$ for $n=1,2,4$ and $T$ the pump period. Light is coupled on-chip by a tapered fiber (TF).}
	\label{fig:SPDC}
\end{figure*}

The setup is shown in Fig. \ref{fig:SPDC} for the case of $N=8$ time bins. The protocol begins with the `pair generation block' in which a pulsed laser at $780$ nm is split into a series of delays of lengths $4T$, $2T$ and $T$, where $T$ corresponds to the desired pump period shown in Fig. \ref{fig:time}. An eight-pulse train is generated that then pumps a nonlinear crystal cut for type-II SPDC, so that a polarizing beam splitter can separate the degenerate signal and idler photons generated at $1560$ nm. The laser power is set such that there is a 5\% chance of multiphoton emission ($\lambda \approx 0.1$). The idler photons are sent to the `heralding decision block,' which consists of a single photon detector (D), data processor (P), bit generator (BG), and decision switch on-chip. The detector, gated by the pump laser, sends time-tagged idler arrival events to the data processor. The processor outputs to a bit generator which modulates the decision switch on-chip, selecting which signal photon will enter the `variable delay circuit' block. The variable delay contains static delays of lengths $4T$, $2T$ and $T$, and 2$\times$2 switches leading in and out of each. These switches allow selection of the coefficients $c_j$. 

The data processing unit required for this protocol will be discussed in detail in Section III and is only outlined here. The processor input is the serial $N$-bit stream from the heralding detector:  1 (0) corresponds to a (no) detection event. The processor writes to a fast bit generator, which modulates at a rate $1/T$ the switch into the variable delay region. The switch is a Mach-Zehnder interferometer (MZI) which can be modulated, e.g., through charge injection \cite{Green:07}. This switch, labeled MZI1 in Fig. \ref{fig:SPDC}, is biased so that if a 0 (1) is received, an incident photon will be rejected from (transmitted to) the variable delay circuit. To drive the variable delay modulators (MZIs 2-5 in Fig. \ref{fig:SPDC}), no data processing is required; simple periodic clock cycles can accomplish this, as shown in Table \ref{tab:logic}.

\begin{table}
	\centering
	\begin{tabular}{| c | c | c | c | c | c |}
		\hline
		Bin & Delay & $\phi_2$ & $\phi_3$ & $\phi_4$ & $\phi_5$ \\
		\hline \hline
		1 & 7T & $\pi$ & 0 & 0 & $\pi$ \\
		2 & 6T & $\pi$ & 0 & $\pi$ & 0 \\
		3 & 5T & $\pi$ & $\pi$ & $\pi$ & $\pi$ \\
		4 & 4T & $\pi$ & $\pi$ & 0 & 0 \\
		5 & 3T & 0 & 0 & 0 & $\pi$ \\
		6 & 2T & 0 & 0 & $\pi$ & 0 \\
		7 & T & 0 & $\pi$ & $\pi$ & $\pi$ \\
		8 & 0 & 0 & $\pi$ & 0 & 0 \\
		\hline
	\end{tabular}
	\caption{The phases $\phi_{2-5}$ in MZ21-5, respectively, required for achieving delays $0$ to $7T$. The values are simply periodic, which allows the modulators to be driven solely by clock signals.}
	\label{tab:logic}
\end{table}

While the high-index contrast of SOI waveguides traditionally results in large linear propagation losses on the order of multiple dB/cm, and switching losses and edge coupling losses on the order of multiple dB, recent advances in fabrication procedures have reduced these figures to allow for scalable AMPP generation at $1560$ nm:

\emph{Switching loss:} The principal sources of loss in our switching elements are free carrier absorption \cite{1073206} from two-photon absorption and mode conversion loss in the MZI directional couplers. By separating the waveguides in directional couplers by more than $150$ nm, mode conversion loss well below $0.1$ dB can be achieved at the expense of device footprint  \cite{Xia:06}. 

Reverse-biased p-i-n modulators can sweep out free carriers on picosecond time scales \cite{Turner-Foster:10}. Assuming the nonlinear refractive index \cite{4866288}, $n_2 = 6\times 10^{-14}$ cm$^2$/W, and a reversed bias figure of merit \cite{Turner-Foster:10} of $0.2$ cm$^{1/2}$ps$^{-1/2}$, a switching efficiency, $\eta_{sw}=0.87$ can be achieved assuming $40$ ps carrier lifetime.  It is possible that nonlinear switch designs will achieve ultra-fast and low-loss switching using the Kerr effect with a pump beam wavelength $>2$ $\upmu$m to reduce two photon absorption \cite{Lee:10}.

\emph{Edge Coupling loss:} SU-8 spot size converters \cite{McNab:03} for coupling large-area fiber modes to SOI waveguides operate with loss of 2-3 dB/facet. Lower losses can be achieved by tapered-fiber coupling to the silicon waveguides, which has been demonstrated for silicon photonic crystal waveguides \cite{2004.IEEE.Srinivasan-Painter.PC_WG_fiber_tapered_coupling} with efficiencies of 94\%. We can include in this loss term the coupling loss into the fiber delay line, and the propagation loss therein. Fiber coupling efficiencies over 90\% have been demonstrated for PPKTP waveguides operating in the near IR \cite{PhysRevLett.93.093601}, and should be applicable to the telecom range.

\emph{Intrinsic linear loss:} Linear loss may be reduced to $\alpha_{lin}=0.1$ dB/cm in low-confinement ridge structures \cite{491567,4866288}. Near-unit efficiency coupling regions from these structures to standard channel waveguides can be fabricated using a standard two-step etch, allowing for low loss delays and high-confinement structures on a single chip \cite{baets_ridge_conversion_2010}. To further reduce loss, we minimize $T$, which is limited ultimately by switching speed in the PIC. On-chip modulators have been demonstrated with switching times shorter than $25$ ps \cite{4375448}, but we assume time bins of $40$ ps.

\section{System Optimization}

We define the AMPP generation efficiency, $\eta$, as the probability that a single photon is emitted at a time interval $NT$ from the AMPP source. $\eta$ depends largely on the efficiency with which signal photons are transmitted through the waveguide switches and delay lines, and the `heralding efficiency' with which the idler photon is detected for heralding. It is useful to consider $\eta$ for the limiting cases of small and large $N$. For small $N$, on-chip loss of signal photons is low because the chip contains only short delays and few switches. $\eta$ is then limited by the heralding efficiency, which is low because of weak pumping and non-unit detection efficiency. As $N$ increases, this heralding efficiency increases. However the number of switches and delay lines increases as well, which can reduce $\eta$ by the loss mechanisms described in the previous section. To find the $N$ that optimizes $\eta$, we now analyze the detection scheme and data processing, considering heralding with (i) a single detector, and (ii) a detector array \footnote{Current InGaAs single photon detector efficiencies remains below 30\%. Efficient up-conversion detection is possible, which is capable of $>90\%$ conversion efficiency and background count rate below $1$ MHz \cite{Albota:04}. Additionally, self-differencing silicon avalanche photodiodes with detection rates up to $1$ GHz are available with detection efficiencies as high as $80\%$ and dark count rates on the order of $1$ kHz \cite{shields_det_patent}. Assuming $85\%$ conversion efficiency and $70\%$ detection efficiency to reduce after-pulsing, a combined detection efficiency of $60\%$ could be achieved for the single detector protocol.}.

(i) For the case of a single heralding detector, one records only the first heralding event, ignoring all consecutive idler photons and dumping all corresponding signal photons. Only the first heralding event is recorded, so that success resulting from time bin $r$, with probability $B(r)$, requires a failure of heralding for all bins $j<r$. Therefore $\eta = \sum_{r=1}^N B(r)$, where the probability of success for the $r^{th}$ bin, 

\begin{equation}
B(r)=(D_{0})^{r-1} \sum_{i=1}^{\infty}H_i\cdot F(r,i).
\end{equation}

$D_0$ denotes the probability that no idler photons are detected in a certain time bin, $H_i$ denotes the probability that $i$ pairs are generated in the given time bin and are then detected, and $F(r,i)$ denotes the probability that $i-1$ of the photons generated in the r$^{th}$ bin are lost in the PIC. We assume Poissonian statistics for ease of calculation. The expressions for these terms are the following:

\begin{equation}
D_{0}=\eta_f\sum_{i=0}^{\infty}e^{-\lambda}\frac{\lambda^{i}}{i!}(1-\eta_{d})^{i}
\end{equation}

\begin{equation}
H_{i}=e^{-\lambda}\frac{\lambda^{i}}{i!}\left(1-(1-\eta_{d})^{i}\right)
\end{equation}

\begin{equation}
F(r,i)=i\left(1-[PIC]^{(r)}\right)^{i-1}[PIC]^{(r)}
\end{equation}

\noindent where $\eta_f$ is the filtering efficiency and $[PIC]^{(r)}$ is the transmission efficiency of the PIC circuit given a photon in bin $r$. Assuming an on-chip coupling efficiency, $\eta_c$, switching efficiency, $\eta_{sw}$ and waveguide transmission efficiency for propagation time $T$, $\alpha_{inc}$, the transmission efficiency becomes

\begin{equation}
[PIC]^{(r)}=\eta_f\eta_{c}(\eta_{sw})^{\lfloor \log_{2}N \rfloor+1}10^{-\alpha_{inc}\cdot(N-r)}.
\end{equation}

For comparison, we also consider the efficiency of a protocol implemented on-chip using a single delay line for time multiplexing, as considered in previous analyses. The model for this system is the same as in Eqs. 3-5, but the on-chip loss becomes

\begin{equation}
[PIC]^{(r)}=\eta_f\eta_{c}(\eta_{sw})^{N-r} 10^{-\alpha_{inc}\cdot(N-r)}.
\end{equation}

\begin{figure}
\begin{center}
	\subfigure[]{\label{fig:edge-a}\includegraphics[scale=0.9]{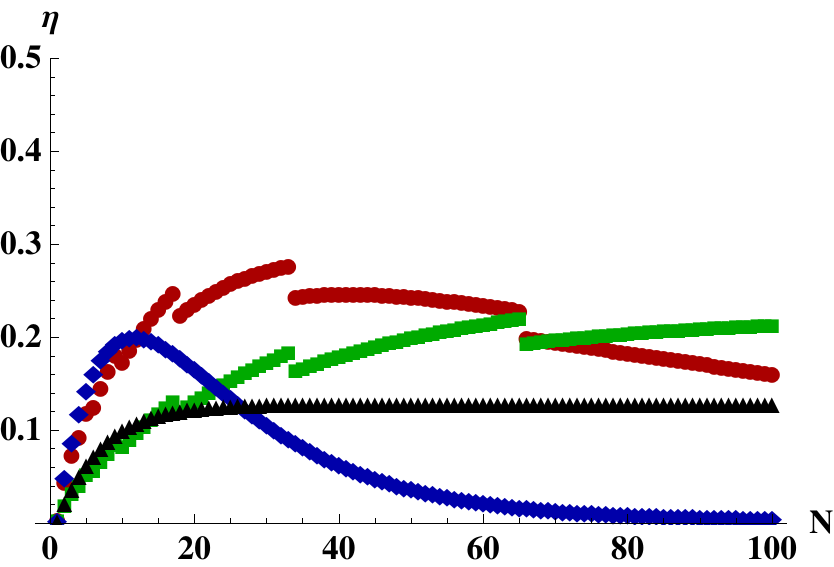}}\\
	\subfigure[]{\label{fig:edge-b}\includegraphics[scale=0.9]{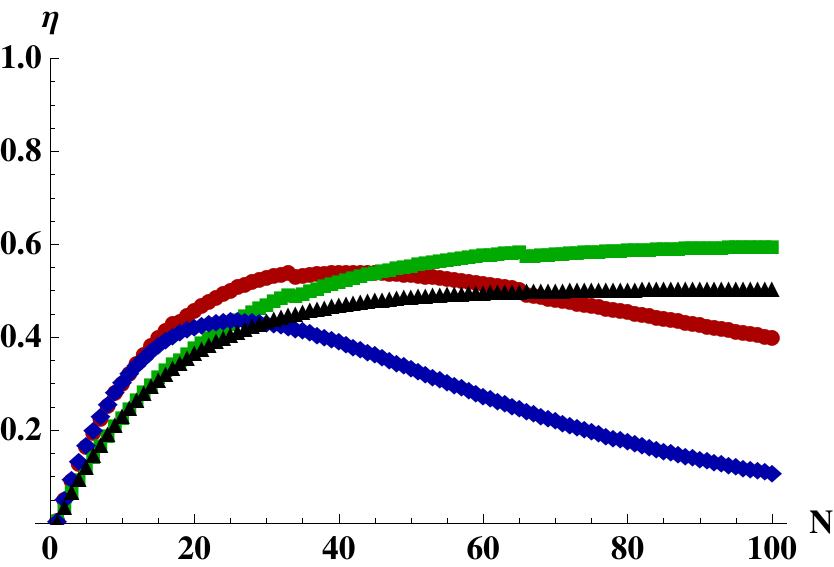}} \\
	\subfigure[]{\label{fig:edge-c}\includegraphics[scale=0.9]{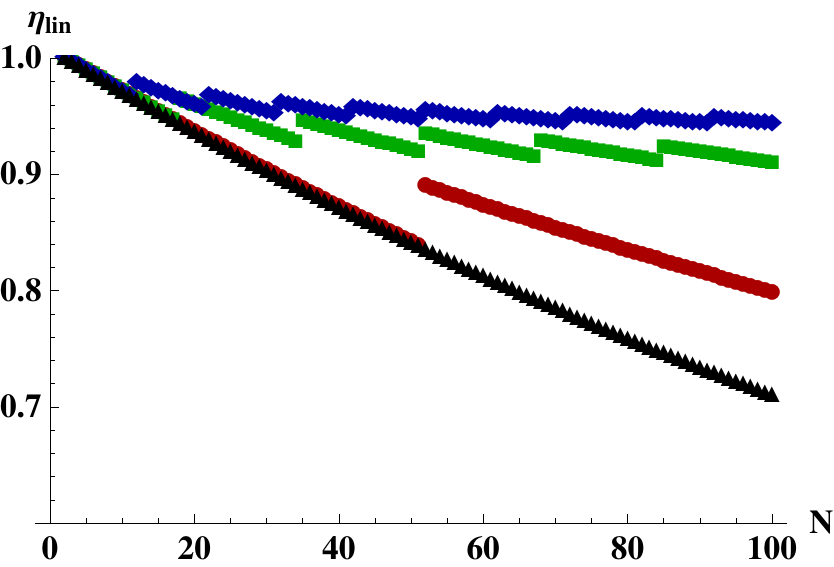}} \\
\end{center}
\caption{(a) The total efficiency of our single detector (red circles) and detector array (green squares) protocol, and previous single-delay-arm time-multiplexing schemes assuming the same single detector (blue diamonds) and detector array (black triangles) protocol, ceteris paribus. This assumes current fabrication procedures, including $\eta_{sw}=0.87$. (b) The same curves as in (a), however assuming $\eta_{sw}=0.98$. (c) $\langle\eta_{lin}\rangle$ for different values of $\lambda$. The red, green and blue curves assume last-photon selection, for $\lambda = 0.02$, $\lambda = 0.06$ and $\lambda = 0.1$, respectively, while the black control curve does not. By last photon selection, the effect of linear loss can be reduced for large N.}
\end{figure}

In Fig. \ref{fig:edge-a} and \ref{fig:edge-b}, we plot the efficiency $\eta$ for the single detector protocol as a function of $N$ assuming parameters given in Table III. We assume switching efficiency of $\eta_{sw}=0.87[0.98]$ for Fig. \ref{fig:edge-a} [\ref{fig:edge-b}]. The AMPP scheme presented in Section II is shown in red, while the previous time multiplexing scheme is shown in blue. 

(ii) Instead of using one detector, detector arrays \cite{migdall_det_array} can be used to detect all idler photons by switching the heralding channel into multiple detection channels. As a result, one can transmit the signal photon that was heralded last and thereby reduce the average delay line loss \cite{1367-2630-6-1-100}, albeit at a loss of net detection efficiency. Fast routing to the individual detectors could be done on chip, so the total detection efficiency, $\eta_d$, includes on- and off-chip coupling and switching losses to $25$ (detector dead time divided by $T$) detectors \footnote{We assume 85\% up-conversion efficiency and a detection efficiency of 80\% which is accompanied by a large after-pulsing probability for the detector array protocol. One can blank a detector after firing for the following period $NT$ to reduce this effect, which is captured in the the factor $24/25$ above.}.

\begin{eqnarray}
\eta_d&=& 0.85\cdot 0.8 \cdot (1/25)[7(\eta_{sw})^4+18(\eta_{sw})^5]\cdot(\eta_c)^2\cdot(24/25) \nonumber \\
&=& 24\%
\end{eqnarray}

The probability of success for the $r^{th}$ bin in this scheme is $B(r)=(D_{0})^{N-r} \sum_{i=1}^{\infty}H_i\cdot F(r,i)$.

For this `detector array' protocol, $\eta$ is plotted in Fig. \ref{fig:edge-a} and \ref{fig:edge-b} as a function of $N$ assuming parameters given in Table III. The AMPP scheme presented in Section II is shown in green, while the previous time multiplexing scheme is shown in black. 

Classical data processing can proceed at a rate $< 1/T$. To account for processing time, the signal photons may be delayed in a low-loss fiber before entering the PIC. Fast field-programmable gate arrays (FPGA) with processing speeds over $1$ GHz are available and suitable for this task; custom circuits could increase speed and decrease optical delay line loss. The FPGA could use the lookup table shown in Table \ref{tab:lut} to realize last photon selection. We estimate that the processing time for $N \approx 50$ should not exceed $1\upmu$s. The propagation loss in the fiber delay arm for this time, assuming $0.2$ dB/km loss, is only $1$\%, and could be lowered with specialized fibers.

\begin{table}
	\centering
	\begin{tabular}{| c | c |}
		\hline
		Input string & Output string \\
		\hline \hline
		10000000 & 10000000 \\
		x1000000 & 01000000 \\
		xx100000 & 00100000 \\
		xxx10000 & 00010000 \\
		xxxx1000 & 00001000 \\
		xxxxx100 & 00000100 \\
		xxxxxx10 & 00000010 \\
		xxxxxxx1 & 00000001 \\
		\hline
	\end{tabular}
	\caption{Lookup table function for last-photon selection. A logical 1 on the output string sets MZI1 in Fig. \ref{fig:SPDC} to pass the corresponding signal photon to the variable delay circuit. $x \in \{0,1\}$. }
	\label{tab:lut}
\end{table}

The advantage of last-photon selection can be examined more closely. The average transmission efficiency through the PIC delay lines, $\braket{\eta_{lin}}$, can be expressed as

\begin{equation}
	\braket{\eta_{lin}}=\left\{ 10^{-\alpha_{inc}(N-i)/10}\right\} _{i=1}^{N}\cdot\left\{ \frac{\prod_{j=1}^{\bar{n}-1}(p-j)}{\sum_{i=\bar{n}}^{N}\prod_{k=1}^{\bar{n}-1}(i-k)}\right\} _{p=1}^{N}.
\end{equation}

\noindent where $\bar{n}=\lceil\lambda N \rceil$. $\braket{\eta_{lin}}$ scales exponentially with the multiplexing parameter, $N$, when $\bar{n}=1$ (black curve in Fig.  \ref{fig:edge-c}), however potentially much more slowly when $\bar{n}>1$ (red circles, green squares and blue diamonds in Fig.  \ref{fig:edge-c}) assuming last photon selection. Jumps in the curve are due to the rounding of $\bar{n}$. Note that the first photon selection required in the single detector protocol trades higher detection efficiency for lower average waveguide transmission, while the last photon selection in by the detector array protocol trades lower detection efficiency for higher average waveguide transmission.
 
Fig. \ref{fig:edge-a} assumes $\eta_{sw}=0.87$ and plots the single detector protocol (red circles) and detector array protocol (green squares) for our scheme, and those for previous time-multiplexing schemes (blue diamonds and black triangles, respectively). Even for this relatively high switching efficiency, the single detector protocol outperforms the detector array protocol. This is reversed in Fig. \ref{fig:edge-b}, where we assume $\eta_{sw}=0.98$. We believe such high switching efficiency could be realized using the nonlinear switches cited earlier \cite{Lee:10}. The crossing point at which the single detector and detector array protocols enable the same maximum $\eta$ occurs for $\eta_{sw} \approx 0.95$.

Fig. \ref{fig:edge-a} assumes realistic fabrication capabilities. A maximum efficiency of 27\% is achieved for $N=31$, or $NT=1.24$ ns using the single detector scheme. The scheme therefore targets photon generation rates of $800$ MHz, achievable with a single heralding detector. The maximum delay length would be $16T$, which corresponds to about $5$ cm in Si. Significantly longer delays have been realized on-chip with low-loss Si waveguides using compact spiral geometries \cite{Dong:10}. Note that the scheme with higher switching efficiencies requires $N=63$, or $NT=2.5$ ns, limiting the generation to $400$ MHz. Slightly higher generation efficiencies could be achieved with larger $N$, however with a slower generation rate. An efficiency of 59\% is achieved for $N=63$.

\begin{table}
	\centering
	\begin{tabular}{| c | c | c |}
		\hline
		Parameter & Value  & Ref.\\
		\hline \hline
		$\eta_{f}$ & $0.99$ & \cite{optigrate} \\
		$\alpha_{lin}$ & $0.1$ dB/cm & \cite{491567,4866288} \\
		$\eta_c$ & $0.84$ & \cite{2004.IEEE.Srinivasan-Painter.PC_WG_fiber_tapered_coupling,PhysRevLett.93.093601} \\
		$\eta_{sw}$ & $0.87$ & \cite{Turner-Foster:10,4866288}\\
		$\eta_{det}$ & $0.7$/$0.8$ & \cite{shields_det_patent}\\
		$\eta_{conv}$ & $0.85$ & \cite{Albota:04}\\		
		\hline
	\end{tabular}
	\caption{Parameters used in the calculation of the single photon generation efficiency plotted in Fig. \ref{fig:edge-a}. We distinguish $\eta_{det}$ from $\eta_d$. The former is the single detector efficiency including up-conversion, whereas the latter is the efficiency of the detection unit for a given operation scheme. For the single detector protocol,  $\eta_{det} = \eta_d$, whereas for the detector array protocol, $\eta_{det} > \eta_d$. $\eta_{det}=0.7$ is used for the single detector scheme where after-pulsing would be a significant problem, and $\eta_{det}=0.8$ is used for the detector array scheme where it's effects can be reduced by blanking detectors after detection. We additionally incorporate the fiber coupling loss and fiber propagation loss of the signal photon into the on chip coupling. We assume 90\% coupling efficiency \cite{PhysRevLett.93.093601}, and 99\% transmission efficiency, as described in Section II and III.}
	\label{tab:eff}
\end{table}

\section{Heralded Bell state generation}

\begin{figure}
\begin{center}
	\subfigure[]{\label{fig:bell-a}\includegraphics[scale=0.35]{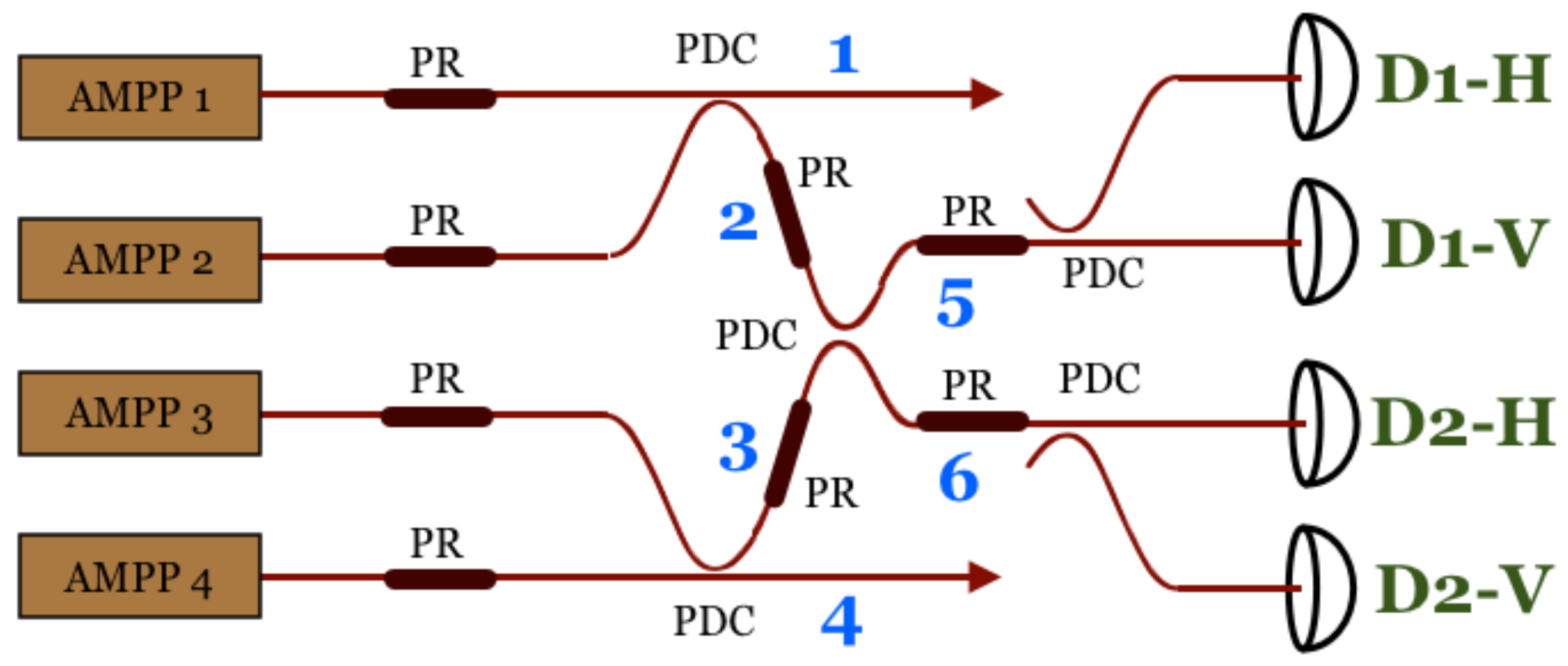}}\\
	\subfigure[]{\label{fig:bell-b}\includegraphics[scale=0.35]{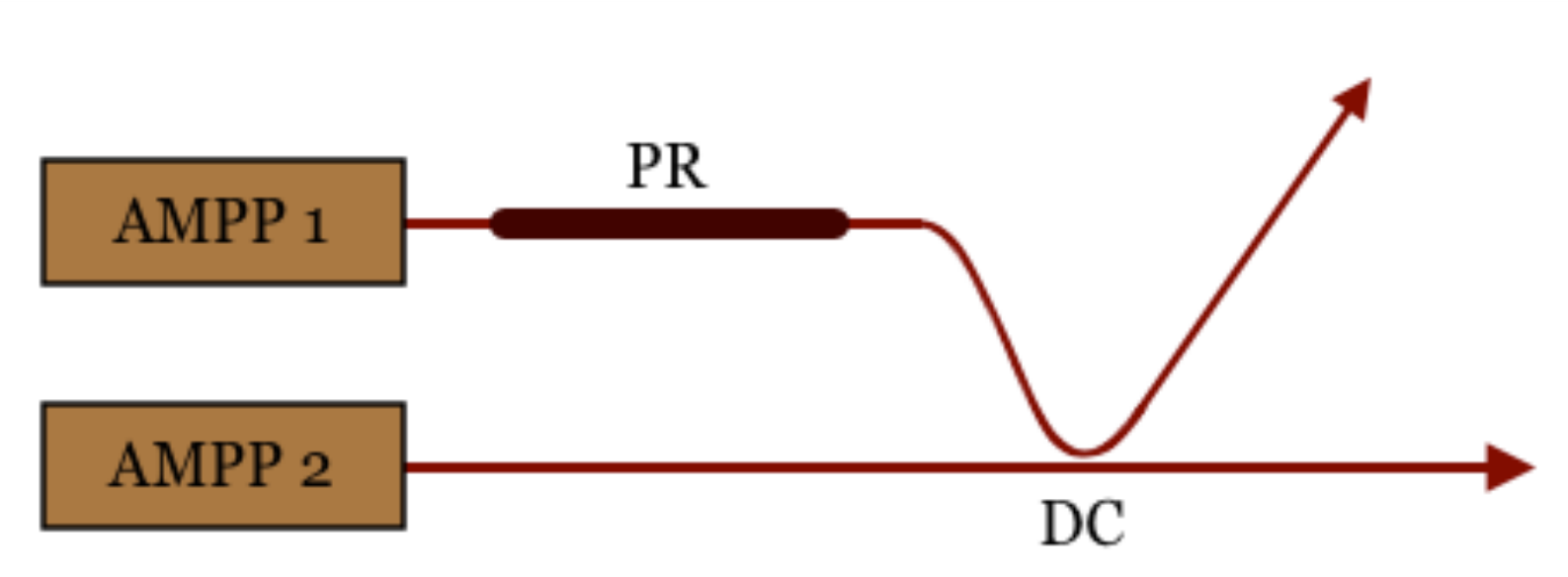}} \\
\end{center}
\caption{(a) Heralded Bell state generation using four AMPP sources. Four photons are combined on two polarizing directional couplers (PDC) after having passed through $\pi/4$ polarization rotators (PR). Ports $1$ and $4$ contain the output state, which is projected onto a Bell state by measurement at the four heralding detectors. These components are analogous to those in bulk optics, and have been realized experimentally on SOI platforms. (b) Entanglement generation from two AMPP sources. Two orthogonally polarized photons interfere on a nonpolarizing directional coupler. Coincidence counting on the two output ports post-selects the state $\ket{HV}-\ket{VH}$.}
\end{figure}

The actively multiplexed single photon source can be used to improve the efficiency of heralded Bell state (HBS)  generation. HBSs are required resources for many quantum information tasks including teleportation for quantum cryptography \cite{PhysRevLett.67.661} and error correction in linear optics quantum computing \cite{klm_2001}. Q. Zhang \emph{et al.} showed that HBSs can be generated with maximum success probability $3/16$ using indistinguishable photons from four single photon sources, together with linear optics and single photon detectors \cite{PhysRevA.77.062316} . This scheme can be implemented on an integrated platform, as shown in Fig. \ref{fig:bell-a}.

Four synchronized AMPP sources produce indistinguishable photons with $\ket{H}$ polarization, which are sent through polarization rotators to produce four H+V states. The upper and lower pairs are collided on polarizing directional couplers (PDC), which transmit $\ket{H}$ and reflect $\ket{V}$. Ports $1$ and $4$ make up the output state, while ports $2$ and $3$ are sent to a third PDC. This coupler operates with a basis rotated by $\pi/4$ with respect to the other PDCs, which can be achieved by placing a $\pi/4$ polarization rotator on each input and output port. A final level of PDCs split ports $5$ and $6$ into four detectors; clicks on detectors D1-H and D2-H or D1-V and D2-V (D1-H and D2-V or D1-V and D2-H) herald the production of the state $\ket{HH}+\ket{VV}$ ($\ket{HV}+\ket{VH}$). This scheme can be compactly implemented on an SOI platform using already demonstrated high efficiency polarization rotators \cite{Watts:05} and polarizing directional couplers \cite{Fukuda:06}. 

The success probability of this scheme is bounded by $\frac{3}{16}\eta^4$, which is only about 4\% considering the high efficiency switch from Table \ref{tab:eff}. However this source can serve as the building block for an active time-multiplexed HBS source. The resulting protocol would follow the same multiplexing principle described in Section III for AMPP generation, however it would take as its input periodic single photons from AMPP sources to generate periodic Bell states.

Alternatively, entangled photons suitable for certain quantum key distribution schemes \cite{PhysRevLett.67.661, PhysRevLett.68.557} can be generated in a simpler way \cite{PhysRevLett.92.037903} using only two AMPP sources, as shown in Fig. \ref{fig:bell-b}. Two synchronized AMPP sources first generate photons with $\ket{H}$ polarization. A $\pi/2$ polarization rotation is applied to one photon, and the photons are combined on a nonpolarizing directional coupler. Coincidence detection on the two output ports post-selects the state $\ket{HV}-\ket{VH}$. In this scheme, the success probability can be as high as $\frac{1}{2}\eta^2$. As with previous demonstrations of these proposals (e.g., \cite{PhysRevA.77.062316}), our system is limited by multiphoton emission, which can be suppressed by reducing the SPDC pump power at the expense of device efficiency.

\section{Conclusion}

We have described a scheme for generating single photon states and Bell states on a PIC. By adopting an active time-multiplexing scheme, it is possible to use only order of $\log_2 N$ switching elements for $N$ single photon generation attempts while maintaining loss scaling with $N$. The loss due to switching is therefore reduced over previous time-multiplexing protocols. A maximum single photon generation efficiency of 27\% can be obtained with a <5\% bound on the conditional multiphoton emission probability, assuming realistic fabrication capabilities. Our scheme is primarily limited by the efficiency of the on-chip switches, which can be assumed to continue to improve rapidly over the next years. If the switching efficiency increases from 87\% to 98\% then the maximum system efficiency will increase from 27\% to 59\%. These sources can additionally form building blocks on-chip that can be combined to generate non-classical quantum optical states, such as Bell states. 

An efficient nonclassical light source on an SOI platform could enable further on-chip integration. Multiple single-photon detectors can already be integrated on-chip \cite{0022-3727-41-9-094010}, and several efforts are now investigating photon pair sources via spontaneous four wave mixing in silicon straight waveguides \cite{Harada:08} and resonators \cite{Chen:11}. With continued development of low-loss structures, self-contained, scalable and reconfigurable photonic quantum information systems could be integrated on a fully CMOS compatible chip.

This work was supported by the DARPA Information in a Photon program, through grant W911NF-10-1-0416 from the Army Research Office.

\bibliographystyle{try.bst}


\clearpage

\end{document}